\documentclass{spie}  

\usepackage{amsmath,amsfonts,amssymb}
\usepackage{graphicx}
\usepackage[colorlinks=true, allcolors=blue]{hyperref}
\usepackage{subcaption}
\usepackage{soul}
\usepackage{upgreek}

\usepackage{xcolor}
\usepackage[T1]{fontenc}

\title{Effects of Proton Irradiation on the Performance of Skipper CCDs}

\author[a]{Brandon Roach}
\author[b]{Brenda A. Cervantes Vergara}
\author[b,d,e]{Santiago Perez}
\author[a,b,c]{Alex Drlica-Wagner}
\author[b]{Juan Estrada}
\author[b]{Abhishek Bakshi}

\affil[a]{Kavli Institute for Cosmological Physics, University of Chicago, Chicago, IL 60637, USA}
\affil[b]{Fermi National Accelerator Laboratory, Batavia, IL 60510, USA}
\affil[c]{Department of Astronomy and Astrophysics, University of Chicago, Chicago, IL 60637, USA}
\affil[d]{Facultad de Ciencias Exactas y Naturales, Departamento de F{\'i}sica, Universidad de Buenos Aires, Buenos Aires, Argentina}
\affil[e]{CONICET --- Universidad de Buenos Aires, Instituto de F{\'i}sica de Buenos Aires (IFIBA),
Buenos Aires, Argentina}

\authorinfo{
BR: roachb@uchicago.edu\\
BCV: bcervant@fnal.gov\\
SP: santiep.137@gmail.com\\
ADW: kadrlica@fnal.gov\\
JE: estrada@fnal.gov\\
AB: bakshi@fnal.gov
}

\begin{document} 
\maketitle

\begin{abstract}
Skipper CCDs are a mature detector technology that has been suggested for future space telescope instruments requiring sub-electron readout noise in the near-ultraviolet to the near-infrared. While modern skipper CCDs inherit from the radiation-tolerant p-channel detectors developed by LBNL,  the effects of high doses of ionizing radiation on skipper CCDs (such as those expected in space) remains largely unmeasured. We report preliminary results on the performance of p-channel skipper CCDs following irradiation with 217-MeV protons at the Northwestern Medicine Proton Center. The total nonionizing energy loss (NIEL) experienced by the detectors exceeds 6 years at the Sun-Earth Lagrange Point 2 (L2). We demonstrate that the skipper amplifier continues to function as expected following this irradiation. Owing to the low readout noise of these detectors, controlled irradiation tests can be used to sensitively characterize the charge transfer inefficiency, dark current, and the density and time constants of charge traps as a function of proton fluence. We conclude with a brief outlook toward future tests of these detectors at other proton and gamma-ray facilities.

\end{abstract}


\section{INTRODUCTION}
\par Charge-coupled devices (CCDs) have become a canonical detector for near-infrared to near-ultraviolet photons. The past several decades of development have sought to optimize a variety of detector parameters, including quantum efficiency, full-well capacity, dark current, and readout noise \cite{Mackay:1986,janesickBook,Lesser:2015}. Of these, the readout noise, $\sigma_r$, is of particular interest for future astronomical CCDs, since it tends to dominate the detector noise budget for faint sources (e.g., exoplanet atmospheres or the most distant galaxies~\cite{Rauscher:2022a, Crill:2022}). Modern astronomical CCDs (e.g., those developed for the Dark Energy Camera \cite{2015AJ....150..150F} or the Dark Energy Spectroscopic Instrument \cite{DESI:2022nlo}) generally have $\sigma_r \gtrsim 2.5$ e$^{-}$ rms/pixel, but for the faintest sources, the true ``holy grail'' is deep sub-electron readout noise that allows the detector to count individual photo-electrons with high precision.
\par The \textit{skipper CCD} was proposed in the early 1990s to overcome the few-electron readout noise of conventional CCDs \cite{10.1117/12.19452,Chandler:1990a}. Unlike conventional CCDs, which only measure the charge in a given pixel once before it is discarded, the skipper architecture incorporates a floating gate amplifier~\cite{Wen:1973a} thereby allowing for repeated nondestructive measurements of the charge within each pixel. If the pixel charge is sampled $N_\text{samp}$ times with single-sample readout noise $\sigma_1$ each time, the final readout noise obtained after averaging all samples is $\sigma_1/\sqrt{N_\text{samp}}$. To reduce the readout time (which could be of order $N_\text{samp}$ times greater than the single-sample case), a readout sequence can be implemented which ``skips'' multiple sampling of irrelevant regions of the detector, hence the name.
\par The sensors described in this work build on the heritage of thick, high-resistivity p-channel CCDs developed by Lawrence Berkeley National Laboratory (LBNL, e.g., ~\cite{Holland:2003,10.1117/12.672393}) and tested for cosmology, particle-physics, and quantum-sensing applications at Fermilab (see, e.g., \cite{Tiffenberg:2017aac,SENSEI:2020dpa,Oscura:2023qik,2024PASP..136d5001V}). In addition to their low readout noise, p-channel skipper CCDs offer several advantages for future space telescope concepts that require single-photon-counting detectors (see, e.g., ~\cite{Rauscher:2022a,Astro2020}). First, compared to other single-photon-sensitive detectors such as the transition edge sensor (TES~\cite{2005cpd..book...63I}), the superconducting nanowire single-photon detector (SNSPD~\cite{10.1063/5.0045990}), or the microwave kinetic induction detector (MKID~\cite{Ulbricht:2021srr}) --- all of which generally require temperatures at or below a few K --- skipper CCDs can operate above 130 K, greatly simplifying the spacecraft thermal and mechanical design. Second, thick p-channel detectors offer excellent quantum efficiency (QE) from wavelengths in the near-UV to near-IR. Third, p-channel detectors have been found to be more radiation-hard than n-channel detectors, a critical attribute for devices that will need to operate consistently for years to decades in deep space (see, e.g., \cite{819155,Marshall:2004a,Dawson:2008a,MORI2013160}).
\par Verifying the radiation-hardness of skipper CCDs is a critical part of the path toward demonstrating technical readiness for space applications. Ionizing radiation (primarily from cosmic-ray protons) induces several kinds of damage in silicon detectors. First, ionizing radiation can lead to charging of the gate dielectric and 
generation of surface trapping states~\cite{janesickBook}. The second kind of radiation damage is nonionizing energy loss (NIEL), typically resulting from the displacement of silicon atoms from their regular lattice positions. The charge traps generated in this process can increase the device's charge-transfer inefficiency (CTI) as well as the dark count rate (see, e.g., ~\cite{osti_7130912,Huhtinen:2002bg,Lindstrom:2002gb,janesickBook}). Finally, there are single-event effects (SEEs) in which a single ionization event can induce a range of undesirable effects in electronics, including single-bit memory flips, transient high-current latchup states, and ruptures between transistor gates leading to device burnout (e.g., ~\cite{see-book,Gaillard2011}).  
%
\par This work describes initial testing to verify the radiation-hardness of p-channel skipper CCDs. In Section~\ref{sec:experiment}, we briefly describe the skipper CCD detectors, the experimental setup at the Northwestern Medicine Proton Center cyclotron, and the Fermilab CCD testing laboratory. In Section~\ref{sec:analysis}, we describe the subsequent testing of these devices at Fermilab and the analysis of the data thereof. We conclude in Sec.~\ref{sec:conclusions} with an outlook toward future work with these devices.
\section{EXPERIMENTAL SETUP}\label{sec:experiment}
\subsection{p-channel CCDs}
The sensors tested in this work are thick (${\sim}700\text{-}\upmu\text{m}$) p-channel skipper CCDs drawing on the LBNL heritage discussed previously. As shown in Table~\ref{tab:detector_list}, most of these sensors are prototype devices for the Oscura~\cite{Oscura:2023qik} and SENSEI~\cite{SENSEI:2020dpa} dark-matter experiments manufactured by two different vendors (hereafter A and B). One sensor was manufactured by a third vendor (hereafter C). All of the skipper CCDs had four readout amplifiers, one at each corner of the sensor. We also irradiated a multi-amplifier sensing 
(MAS~\cite{Holland:2023, Botti:2024}) CCD from vendor C, though testing of this detector is outside the scope of the present study.
\begin{figure}
    \centering

    \includegraphics[width=0.7\textwidth]{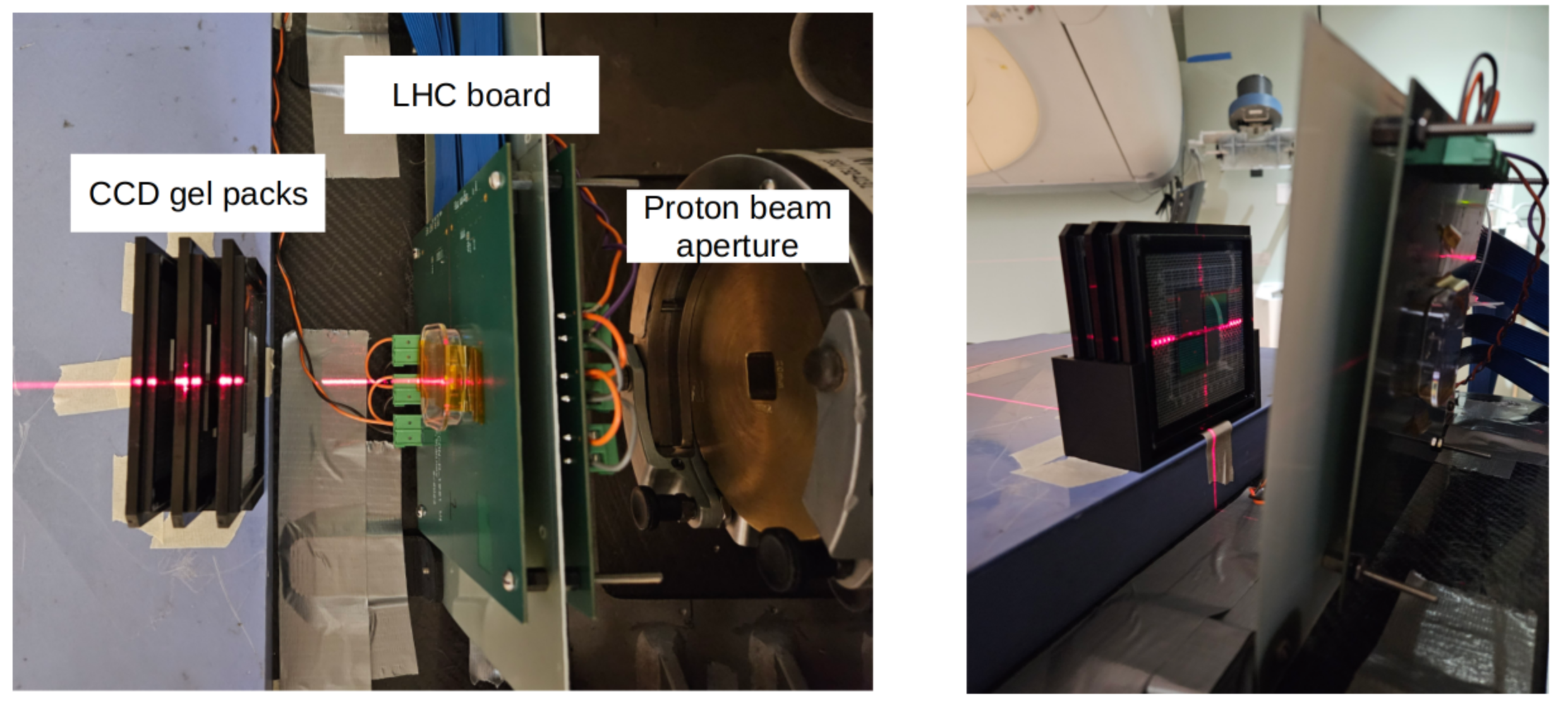}

    \caption{(\textbf{Left}) Top-down view of the setup at the accelerator, showing (from left to right) the three CCD gel packs, Large-Hadron-Collider (LHC) group electronics board, and the proton beam aperture. The distance between the first gel pack and the beam aperture is 20 cm. The red lines are the alignment lasers used for targeting the beam. (\textbf{Right}) Front view of the setup, showing the four sensors in the first gel pack.}
    \label{fig:layout}
\end{figure}
\begin{figure}
    \centering
    \includegraphics[width=0.8\textwidth]{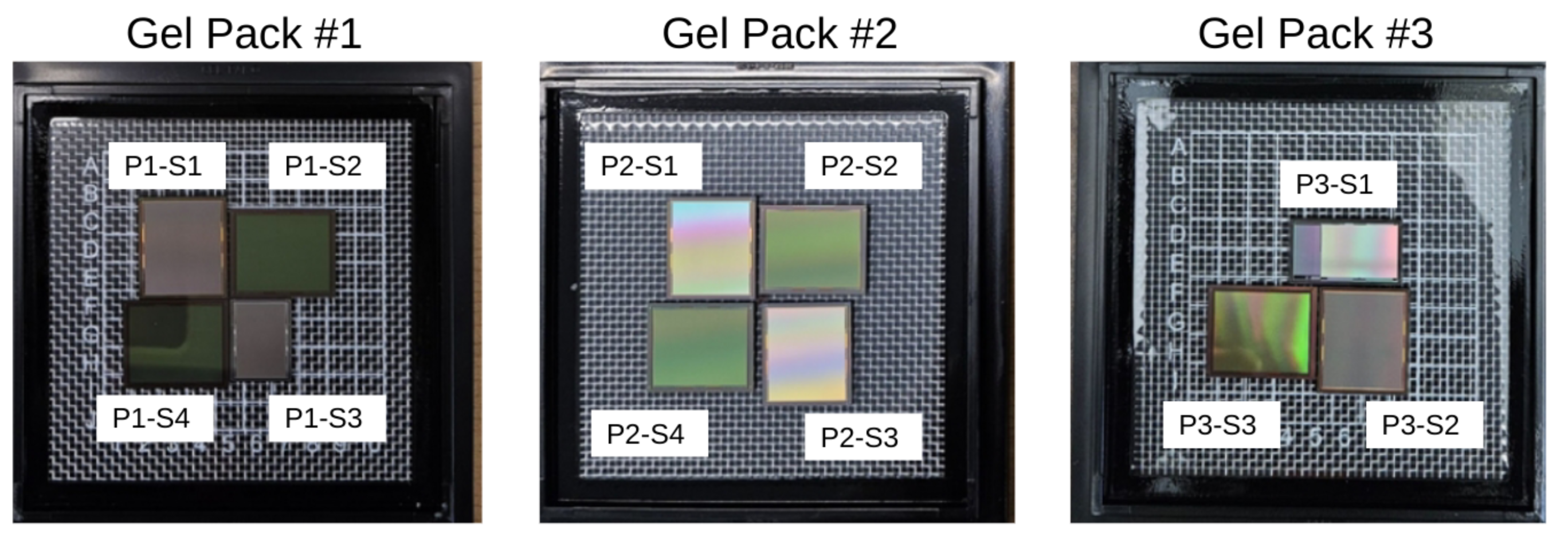}
    \caption{Layout of the sensors within the three gel packs irradiated. Sensors are listed in Table~\ref{tab:detector_list}. The center of the square proton beam was aligned with the center of the gel packs.}
    \label{fig:gel_pack_layout}
\end{figure}
\begin{figure}
    \centering
    \includegraphics[width=0.75\textwidth]{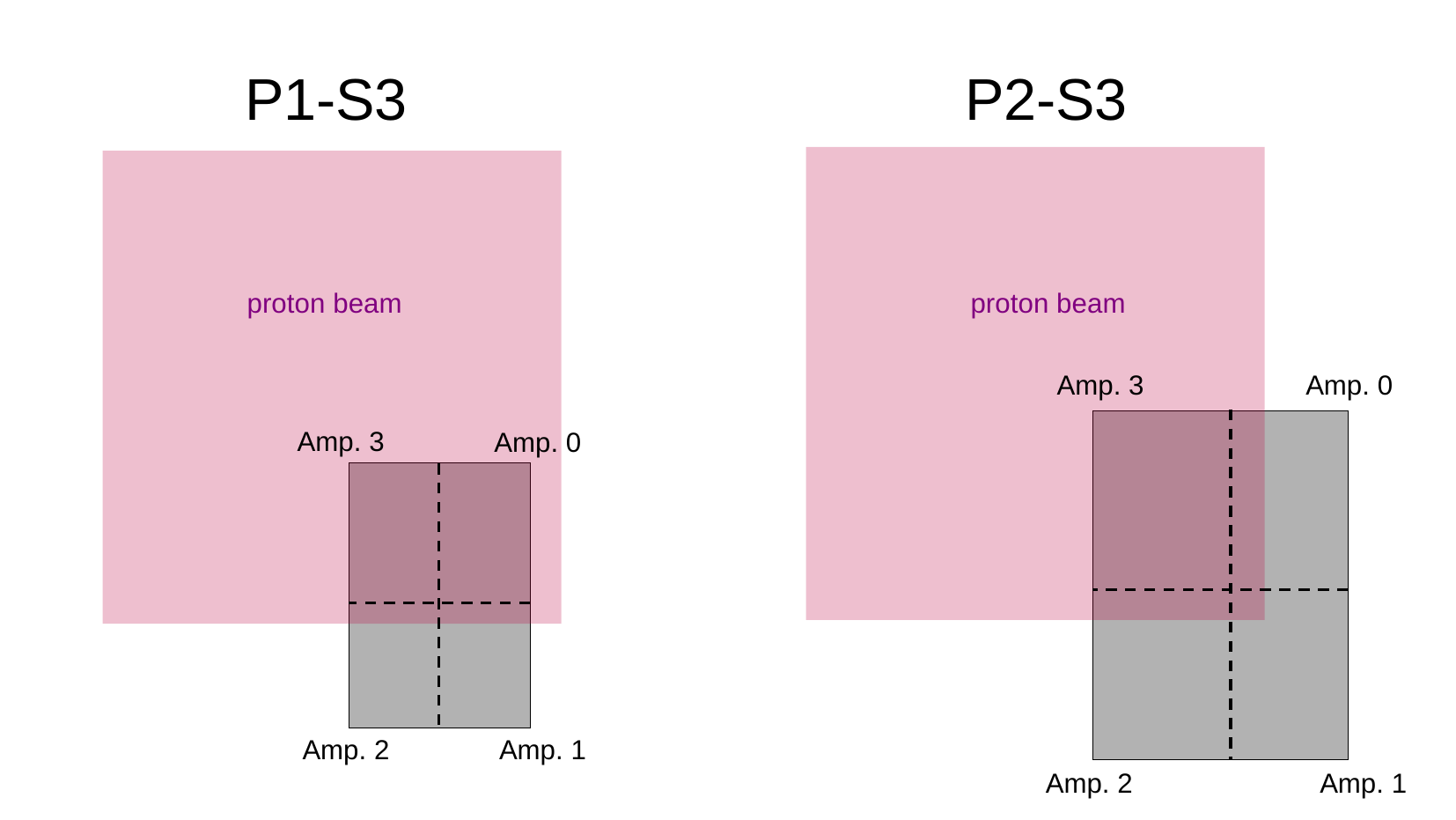}
    \caption{Sketch showing the approximate areas of sensors P1-S3 (left) and P2-S3 (right) irradiated by the proton beam (purple square), as traced by the density of charge traps revealed by pocket-pumping (see Section 3.2).}
    \label{fig:irrad_sketch_ccd_area}
\end{figure}

\subsection{Northwestern Medicine Proton Center}
The Northwestern Medicine Proton Center in Warrenville, IL, hosts a 217-MeV proton cyclotron used for cancer treatment. At a distance of 10 cm from the exit aperture and with no additional collimation, the beam profile takes the form of a square with area $(1.8\text{\,cm})^2$. Within this region, the beam intensity is uniform to within a few percent. The beam intensity is calibrated assuming a distance of 10 cm from the exit aperture. Since our detectors were 20 cm from the beam aperture, the beam intensity was reduced by ${\sim}10\%$ from the nominal value. The proton fluence (cm$^{-2}$) through the sensors can be controlled by varying the beam current (in the range of ${\sim}$10--100 nA) and/or irradiation time. An ionization chamber is used to measure the total dose delivered. Finally, behind the irradiated devices is a stack of Solid Water~\cite{GARGETT202043} blocks to stop the beam. 
\par Initial irradiation tests were performed in August 2023. These tests were primarily focused on the effects of bulk displacement damage arising from NIEL. The \textit{NIEL scaling hypothesis} (see, e.g., ~\cite{Huhtinen:2002bg,Lindstrom:2002gb}~and references therein) posits that the bulk damage is directly proportional to the displacement damage cross section, $D_i(E)$, and to the particle fluence, $\Phi_i$ (here $i$ labels the type of incident particle). To compare the effects of different particles and energies, we use the hardness factor (e.g.,~\cite{Lindstrom:2002gb})
\begin{equation}
    \kappa_i(E) = \frac{D_i(E)}{D_n(1\text{\,MeV})},
\end{equation}
normalizing to the cross section for 1-MeV neutrons (approximately 95 keV barn in silicon~\cite{ASTM}). For the 217-MeV protons at Warrenville, $\kappa_p \approx 1$, whereas $\kappa_p \approx 3.5$ for the 12.5-MeV protons used in previous work~\cite{Dawson:2008a}~. Thus, we required approximately 3.5 times greater proton fluence to reach the same damage level as achieved in previous work~\cite{Dawson:2008a}~. The minimum fluence tested corresponds to $\gtrsim$6 years expected radiation dose for a satellite at the Earth-Sun L2 point using the fluences quoted in ~\cite{Dawson:2008a}.

\par For the tests in August 2023, we irradiated the sensors at ambient conditions (room temperature and pressure). To hold the sensors in the beam, as shown in Figures~\ref{fig:layout}--\ref{fig:gel_pack_layout}, we stuck the sensors to  adhesive gel carrier packs. We also arranged the sensors such that at least one quadrant of each sensor (i.e., at least one amplifier and one-quarter of the active area) was out of the beam path to serve as a control when assessing radiation damage. The approximate area of each sensor exposed to the proton beam is shown in Figure~\ref{fig:irrad_sketch_ccd_area}. Each gel pack was irradiated for several minutes until the desired dose was reached. Following irradiation, each gel pack was covered to protect from external contamination and temporarily moved to an adjacent room to wait for the induced radioactivity to return to safe levels. At the end of the day, all three gel packs were returned to room-temperature dry storage at Fermilab to await packaging and testing.

These irradiation tests were conducted at the same time as another Fermilab group was testing the radiation hardness of readout electronics for the CMS experiment at the Large Hadron Collider (LHC). Their PCB was placed between the beam aperture and our sensors (Figure~\ref{fig:layout}), but at these energies, the attenuation of the beam was minimal.
\subsection{Fermilab testing}
The packaging and testing of the CCD sensors at Fermilab proceeded similarly to those used for Oscura detector prototypes (e.g.,~\cite{Oscura:2023qik}). First, each sensor was epoxied to a silicon substrate and wirebonded to a kapton flex cable which provided the electrical connections. Each sensor was then mounted into a copper tray and placed inside the vacuum chamber used for characterizing the Oscura sensors. We used a single low-threshold acquisition (LTA~\cite{Cancelo:2020egx}) board for CCD control and readout. A closed-cycle cryocooler, resistive heater, and temperature sensor were coupled to the copper tray, allowing us to maintain a 150-K operating temperature for the CCDs with a Lakeshore temperature controller. A small violet LED was also placed inside the vacuum chamber to illuminate each CCD for the pocket-pumping tests described in the next section.  
\begin{table}[t]
    \centering
    \begin{tabular}{c|c|c|l}
        Detector & Position & Dose ($10^{10}\text{\,cm}^{-2}$) & {\hfill Comment \hfill} \\
        \hline
        P1-S1 & Upper left & 1.2 & Manufacturer A, Oscura first batch \\
        P1-S2 & Upper right & 1.2 & Manufacturer B, Oscura first batch \\
        P1-S3 & Lower right & 1.2 & Manufacturer B, Oscura first batch \\
        P1-S4 & Lower left & 1.2 & Manufacturer C, SENSEI small die \\
         & & \\
         P2-S1 & Upper left & 8.4 & Manufacturer A, Oscura first batch\\
         P2-S2 & Upper right & 8.4 & Manufacturer B, Oscura first batch \\
         P2-S3 & Lower right & 8.4 & Manufacturer B, Oscura first batch \\
         P2-S4 & Lower left & 8.4 & Manufacturer A, Oscura first batch \\
         & & \\
        P3-S1 & Top & 31.9 & Manufacturer C, 8-amplifier MAS \\
         P3-S2 & Lower right & 31.9 & Manufacturer B, Oscura first batch \\
         P3-S3 & Lower left & 31.9 & Manufacturer A, Oscura first batch \\
    \end{tabular}
    \vspace{1em}
    \caption{Detector inventory for each gel pack tested. Naming format is (gel pack number)-(sensor number). All detectors are 4-amplifier skipper CCDs except for P3-S1, which is an 8-amplifier multi-amplifier sensing (MAS) CCD \cite{Holland:2023}. See Figures~\ref{fig:layout} and \ref{fig:gel_pack_layout} for layout. The 217-MeV proton dose has been corrected for the 20-cm distance of the detectors from the beam aperture.}
    \label{tab:detector_list}
\end{table}

\section{DATA ANALYSIS}\label{sec:analysis}
This section describes the data analysis for two sensors, labeled P1-S3 and P2-S3 in Table~\ref{tab:detector_list}, performed within two weeks after irradiation. We seek to verify that the detectors still function after proton bombardment, and to quantify the degree of radiation damage as a function of proton dose. We note that these two sensors are different sizes: P1-S3 is $1020 \times 682$ pixels, and P2-S3 is $1278\times 1058$ pixels, though both have $(15\,\upmu\text{m})^2$ pixels.

\subsection{Skipper amplifier performance}
One of the main goals of these initial tests was to demonstrate that these skipper CCDs would still be able to perform repeated nondestructive charge measurements following irradiation, and that the noise performance would follow the expected $1/\sqrt{N_\text{samp}}$ dependence ($N_\text{samp}$ is the number of samples).. We note that one of the four amplifiers on each detector did not function normally during these tests: amplifier 2 on P1-S3 and amplifier 3 on P2-S3. For the former, the images contain no charge; for the latter, the images contain several cosmic-ray tracks, but there is substantial charge streaking during readout. However, we note that these amplifiers functioned normally in the configuration used for the pocket-pumping tests described in the next section. We have previously observed similar behavior on non-irradiated sensors; thus, we attribute this behavior to inherent issues with the sensors, rather than a radiation-induced effect.
\par
\begin{figure}[h]
    \begin{subfigure}{0.42\textwidth}
        \includegraphics[scale=0.4]{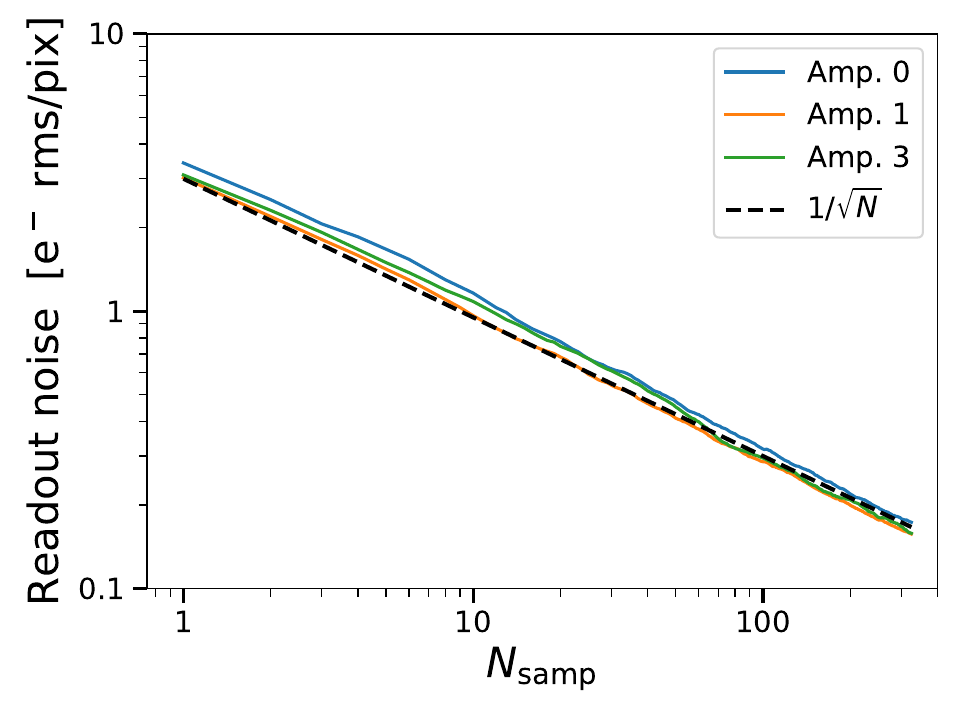}
        \label{fig:noise_vs_N_I01}
    \end{subfigure} \hspace{0.075\textwidth}
    \begin{subfigure}{0.42\textwidth}
        \includegraphics[scale=0.4]{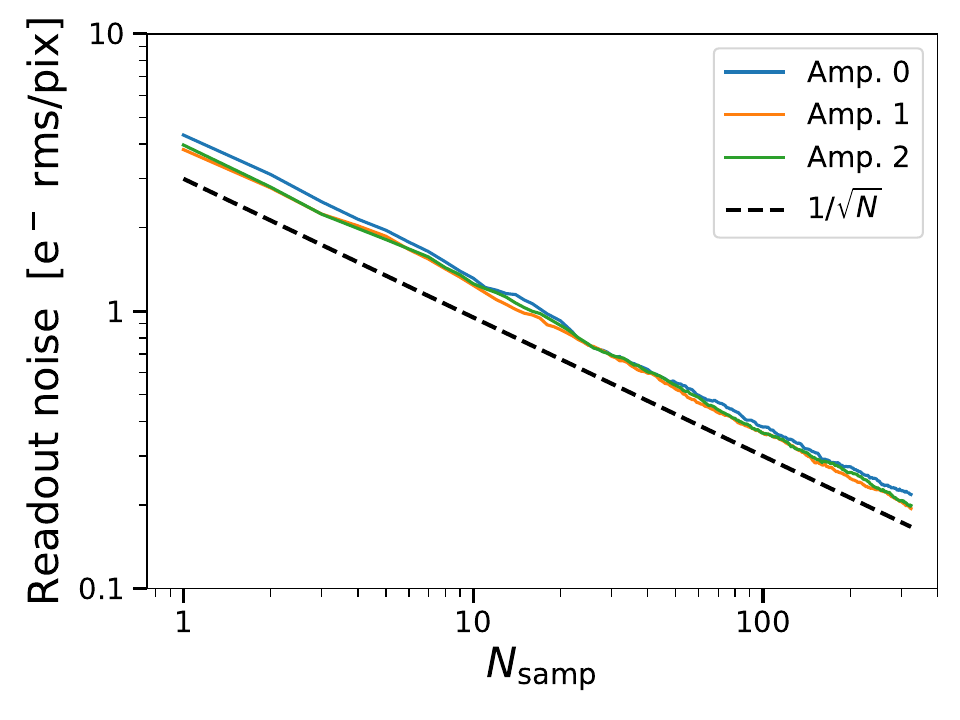}
        \label{fig:noise_vs_N_I02}
    \end{subfigure}
\caption{Scaling of detector readout noise versus $N_\text{samp}$ for sensor P1-S3 (left) and P2-S3 (right). The black dashed lines show the expected scaling for a detector with a single-sample readout noise of 3 e$^{-}$ rms/pix.}
\label{fig:noise_vs_N}
\end{figure}

 In Figure~\ref{fig:noise_vs_N} we show the results from the three working amplifiers on sensors P1-S3 and P2-S3, obtained from the overscan region of zero-exposure bias images. The data from each amplifier consists of a list of 324 images with dimensions $50\times 650$ pixels (including the 50-column-wide overscan), with each image representing a single nondestructive sample of the pixel charge. We construct cumulative running-average images as a function of $N_\text{samp}$ and fit the pixel charge distributions  with Gaussian peaks to determine the readout noise and amplifier gain (ADU/e$^-$). The three working amplifiers on P1-S3 (P2-S3) have single-sample readout noise ${\sim}$3--3.5 e$^{-}$ rms/pix (${\sim}$4--4.5 e$^{-}$ rms/pix). These are consistent with previous measurements of non-irradiated skipper CCDs for the SENSEI and Oscura experiments~\cite{SENSEI:2020dpa,Oscura:2023qik}. All of these  amplifiers also follow the expected $1/\sqrt{N_\text{samp}}$ behavior up to at least 324 samples, where the readout noise is $\lesssim 0.2\text{--}0.3\text{\,e}^{-}\text{\,rms/pix}$.  In any case, this work demonstrates that the multiple nondestructive sampling capability of skipper CCDs is preserved after proton irradiation, which is a crucial result for space applications.
\begin{table}[t]
    \centering
    \begin{tabular}{c|c|c}
        Sensor & Trap density [irrad.] & Trap density [non-irrad.]\\
        \hline
        P1-S3 & $2\times 10^{-2}$ (amps. 0, 3)& $1\times 10^{-3}$ (amps. 1, 2)\\
        P2-S3 & $4\times 10^{-2}$ (amp. 3) & $3\times 10^{-3}$ (amps. 1, 2)
    \end{tabular}
    \caption{Density of traps (in traps/pixel of active area) from the irradiated and non-irradiated sections of sensors P1-S3 and P2-S3. The parentheses indicate which amplifiers' data were used in the trap density calculation.}
    \label{tab:trap_density}
\end{table}
\subsection{Pocket pumping}
To study the behavior of the charge traps in these CCDs, we used ``pocket pumping'' (e.g., ~\cite{Blouke1988, janesickBook}). In this technique, the CCD is uniformly illuminated with light (in this case, from a violet LED) to build up ${\gtrsim}10^3$ electrons per pixel, and the clocking sequence is manipulated to repeatedly move the charge back and forth between pixel phases. As the charge packet moves over a trap, some of the charge may be captured and subsequently re-emitted during another step of the clock cycle (in this case, the duration of each step in the clock cycle,  $t_\text{ph}$, was a constant 667$\,\upmu\text{s}$). If this process is repeated for $N$ times before reading out the device, a charge dipole will be built up in adjacent image pixels, showing the location of the trap. 
\par
\begin{figure}[h]
    \centering
    \begin{subfigure}{0.48\textwidth}
        \includegraphics[scale=0.5]{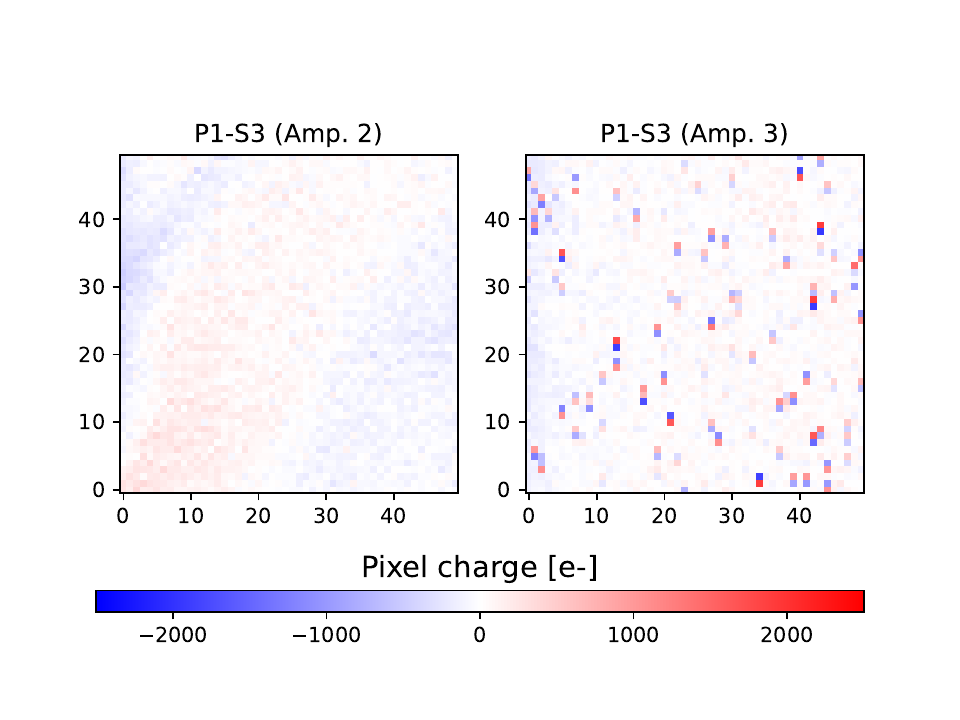}
        \label{fig:trap_images_I01}
    \end{subfigure}
    \hspace{0.3cm}
    \begin{subfigure}{0.48\textwidth}        \includegraphics[scale=0.5]{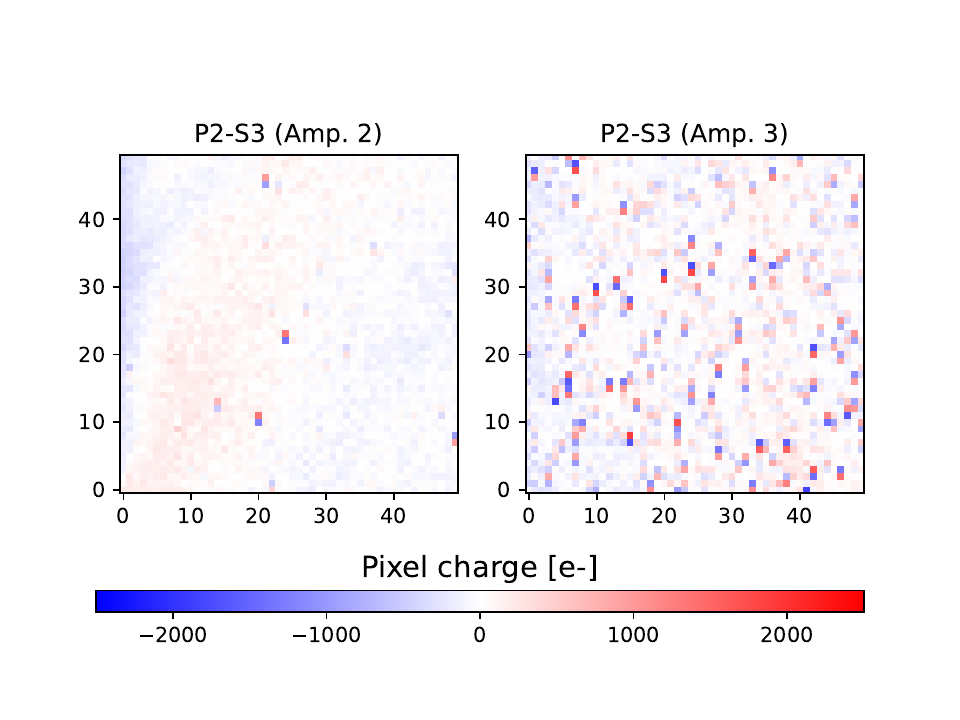}
        \label{fig:trap_images_I02}
    \end{subfigure}
    \caption{$50\times 50$ pixel cutouts of data from amplifiers 2 and 3 of sensors P1-S3 (left) and P2-S3 (right) obtained after pocket pumping for $2\times 10^4$ cycles at 150 K, with $t_\text{ph} = 667\,\upmu\\text{s}$. Each image was taken with $N_\text{samp} = 10$, giving a readout noise ${\sim} 1\text{\,e}^{-}$ rms/pix. The increase in trap density between amplifier 2 (non-irradiated) and amplifier 3 (irradiated) is clearly visible, as is the greater density of traps on amplifier 3 of P2-S3, using $t_\text{ph} = 667\,\upmu\text{s}$.}
    \label{fig:trap_images_both}
\end{figure}

 Our detailed analysis of the traps generated by proton irradiation (including time constants and trap energies) is ongoing; here, we present some general remarks on sensors P1-S3 and P2-S3. Based on the positions of the sensors in the gel packs, we expected amplifiers 0 and 3 would be  within the beam area, whereas amplifiers 1 and 2 would be  outside the beam area. Thus, the quadrants of the detectors read by amplifiers 0 and 3 would be expected to have a higher trap density than amplifiers 1 and 2. (We emphasize again that all four amplifiers on both sensors functioned normally in the configuration used for pocket-pumping tests.) As shown in Figure~\ref{fig:trap_images_both} and Table~\ref{tab:trap_density}, this is indeed what we observe. To compute the trap density in sensor P1-S3, we take the average of the densities from amplifiers 0 and 3, since both of those quadrants were similarly irradiated. For sensor P2-S3, we only use the data from amplifier 3, since the other quadrants were largely outside the beam path. (For both sensors, we use the sections of the images from amplifiers 1 and 2 that were farthest from the proton beam path to calculate the trap density for the non-irradiated regions.) One curious result is that the trap density does increase between P1-S3 and P2-S3, but by only a factor of ${\sim}2$ versus the factor of ${\sim}7$ difference in proton fluence experienced by these sensors. We believe this may be a result of the code we are using to extract the dipoles becoming less efficient when the trap density is very high. In future irradiation tests, we plan to expose the sensors to much lower proton doses to more efficiently extract and characterize the traps.

\section{CONCLUSIONS AND OUTLOOK}\label{sec:conclusions}
\par In this work, we described initial tests of p-channel skipper CCDs irradiated with high-energy protons. We packaged and tested two of the sensors at Fermilab within two weeks after irradiation, and found that three of the four amplifiers on each detector functioned well. We do not believe the issue with the other amplifiers was a result of irradiation. These tests also demonstrated for the first time that the repeated nondestructive sampling capability of the CCD was preserved following irradiation. Further tests of the detectors listed in Table~\ref{tab:detector_list} are ongoing, and will be used to determine the effects of higher proton fluences on trap behavior, dark current, CTI, etc.
\par
\begin{figure}[t]
    \centering
    \includegraphics[width=0.65\textwidth]{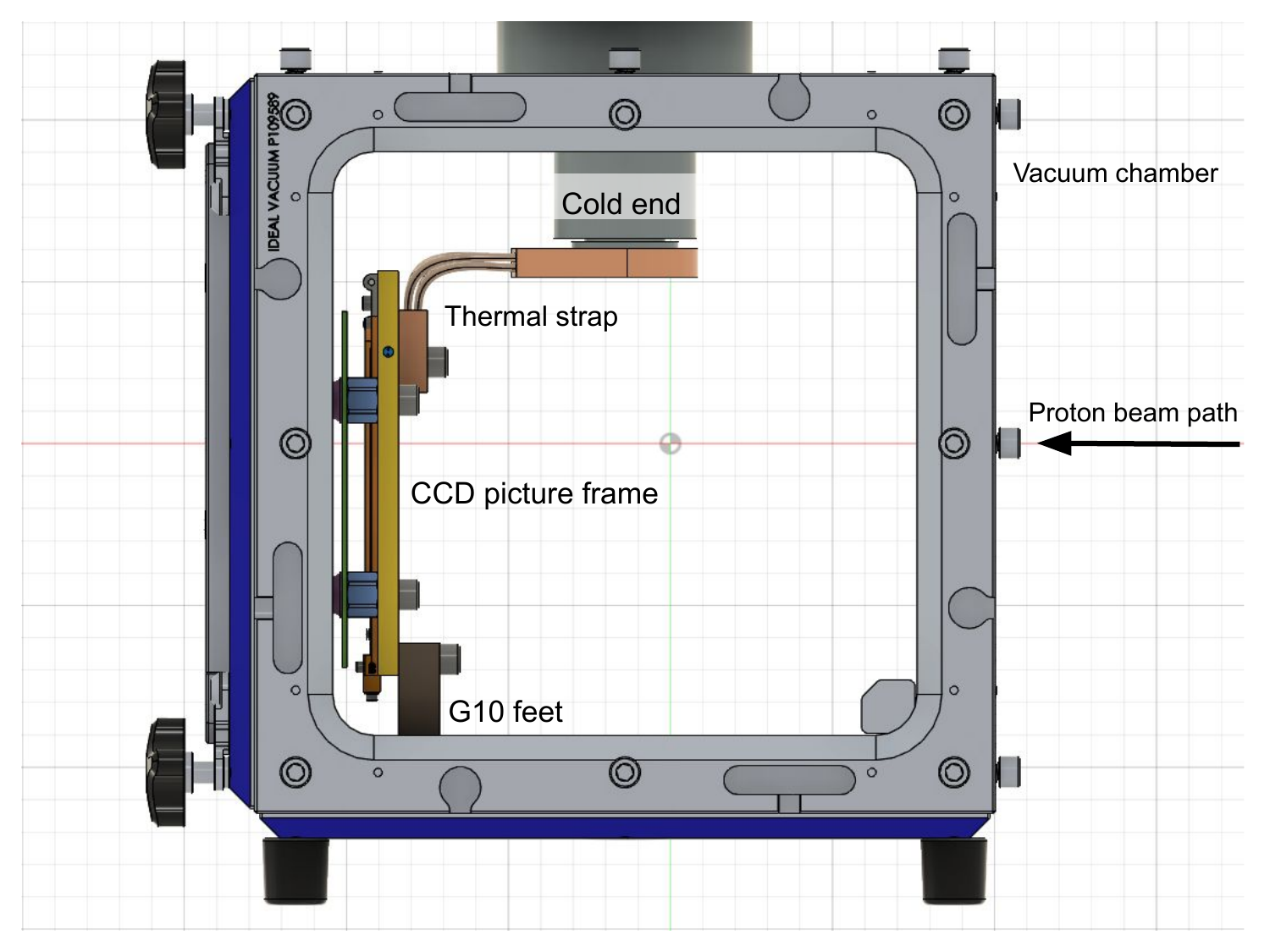}
    \caption{Rendering of the Fermilab vacuum chamber designed for future irradiation tests of skipper CCDs. The proton beam path is indicated by the horizontal arrow and red line.}
    \label{fig:radiation_cube}
\end{figure}

 For future irradiation tests, we plan to make several modifications to the setup to achieve a more realistic operating environment (thermal and vacuum) for the CCDs during irradiation. We are in the final stages of constructing a vacuum chamber to contain one CCD at a time for proton irradiation (e.g., at the Northwestern Medicine Proton Center) and subsequent testing at Fermilab. A rendering is shown in Figure~\ref{fig:radiation_cube}. This chamber is based on the Ideal Vacuum modular vacuum chamber used by the CCD group at Fermilab, though with several modifications to reduce the likelihood of chamber materials becoming activated by the proton beam. First, the proton beam will enter and exit the chamber through 120-$\upmu\\text{m}$-thick titanium foil windows, which will maintain the vacuum while reducing the mass in the path of the beam. Furthermore, if the titanium foils do become irradiated, their disposal and replacement is much simpler than large pieces of aluminum. Second, a new ``picture frame'' (see, e.g., \cite{2006SPIE.6276E..08D}) to hold the sensor has been designed, and will incorporate an aluminum nitride (rather than copper) support board electrically and thermally isolated from the chamber floor by G10 feet. Finally, since the cold end of the cryocooler will need to be offset from the path of the proton beam, we will thermally couple the CCD to the cold end via a copper strap routed away from the beam path. We anticipate this system will be ready for detector testing by mid-summer 2024. 

\acknowledgments 
  We are grateful to the staff of the Northwestern Medicine Proton Center for allowing us to use their beamline. We also thank Jim Hirschauer and the other members of the Fermilab CMS group for their assistance with planning our irradiation experiment,  our Fermilab CCD colleagues for their assistance with detector processing and packaging, and Bernie Rauscher and Steve Holland for their helpful comments and suggestions. The fully depleted skipper CCD was developed at Lawrence Berkeley National Laboratory, as were the designs described in this work.
This manuscript has been authored by Fermi Research Alliance, LLC under Contract No. DE-AC02-07CH11359 with the U.S. Department of Energy, Office of Science, Office of High Energy Physics. The U.S. Government retains and the publisher, by accepting the article for publication, acknowledges that the U.S. Government retains a non-exclusive, paid-up, irrevocable, world-wide license to publish or reproduce the published form of this manuscript, or allow others to do so, for U.S. Government purposes. This work was partially supported by the Fermilab Laboratory Directed Research and Development program (L2019.011 and L2022.053), NASA APRA award No. 80NSSC22K1411 and a grant from the Heising-Simons Foundation (\#2023-4611). BR is supported by a KICP Fellowship at the University of Chicago.

\bibliography{report} 
\bibliographystyle{spiebib} 

\end{document}